# Undoped and Eu, Na co-doped LiCaAlF$_6$ scintillation crystals: paramagnetic centers, charge trapping and energy transfer properties

M. Buryi[1*], V. Laguta[1], V, Babin[1], J. Pejchal[1], M. Nikl[1], A. Yoshikawa[2]

[1]*Institute of Physics CAS in Prague, Cukrovarnicka 10/112, Prague, Czechia*
[2]*Institute for Materials Research, Tohoku University, 2-1-1 Katahira, Sendai 980-8577, Japan*

**Abstract**

Single crystals of LiCaAlF$_6$ undoped and Eu, Na co-doped were studied by electron paramagnetic resonance, radioluminescence and thermally stimulated luminescence techniques applied in a correlated manner. The undoped samples exposed to X-ray irradiation exhibited two hole-like charge trapping centers creation, the molecular ions of the form: ClF$^-$ and F$_2^-$ – F$_2^-$ dimer. Their trap depths and frequency factors were determined as follows: $E_t(\text{ClF}^-) = 1.7 \pm 0.1$ eV and $E_t(\text{F}_2^- - \text{F}_2^-) = 1.1 \pm 0.1$ eV for trap depths and $f_0(\text{ClF}^-) = f_0(\text{F}_2^- - \text{F}_2^-) \sim 10^{13}$ s$^{-1}$ for frequency factor, respectively. It was found that the europium preferable charge state is 2+ in the LiCaAlF$_6$:Eu,Na samples, however, some amount of the Eu$^{3+}$ is also present. Moreover, there were two Eu$^{2+}$ centers: the dominating $\text{Eu}_1^{2+} = \text{Eu}_{\text{Ca}}^{2+}$ and the low-content $\text{Eu}_2^{2+} = \text{Eu}_{\text{Li}}^{2+}$. The amount of the latter is easily governed by the sodium admixture while the former is insensitive to the Na co-doping. Eu and Na co-doping affected the defects distribution and incorporation in the LiCaAlF$_6$ host.

## 1. Introduction

Medical applications such as radiography and tomography, disease diagnostics and therapy among great amount of other ionizing radiation applications require the most efficient scintillating materials of the highest possible energy and space resolution. Non-destructive neutron radiography and tomography are particularly sensitive diagnostics methods benefitting from extended penetrating abilities of neutrons as compared to X-rays. Besides, neutron intensities are perceptive enough to detect water and hydrogen-containing compounds. Activation of the LiCaAlF$_6$ (LiCAF) crystals with europium ions gives way to effective neutron scintillators (detectors) engineering [1], non-hygroscopic and of high transparency. Remarkably, co-doping of LiCAF with Eu and Na resulted in increased light yield [2]. These crystals are fabricated to replace the $^3$He-based counters because of the thinning $^3$He supplies [3]. LiCaAlF$_6$ is extremely good due to the following features: (i) naturally occurred $^6$Li isotope (the material can also be intentionally enriched with that one). The $^6$Li(n,α)$^3$H reaction leads to high cross section for thermal neutron capture [4]; (ii) low effective atomic number, Zeff = 14, and the low density of 2.98

g/cm$^3$ are expected to create conditions for effective background γ-rays suppression [5]; (iii) non-hygroscopic; (iv) big volume crystals availability. The Eu-doped LiCAF crystals exhibited high light yield (LY), about 30000 ph/n [1]. Co-doped with Na it exhibits the light yield of 40000 ph/n [2]. Never-ending search for further improvement leads to several very recent works dedicated to solid solutions of the LiCaAlF$_6$ and LiSrAlF$_6$ [6]. The LY of the resultant solid solutions was within the 22000-30000 ph/n. The LiCaAlF$_6$ doped with Nd demonstrated dose dependence of the thermally stimulated luminescence (TSL) glow curve so it is expected to have a potential to be realized in high-dose irradiation measurements [7] as well as the LiCaAlF$_6$:Tb [8] revealing similar properties. The LiCaAlF$_6$ was even tried in the micro/nano particles form prepared by laser ablation [9]. The corresponding photoluminescence temporal profiles exhibited fast-decay alongside the slow-decay components in the bulk crystal. Smaller particles demonstrated shorter decay times than larger ones. For example, the fast-decay component of the particles having averaged diameter smaller than 0.36 μm was about 40 ns. It is shorter by at least one order of magnitude as compared to the bulk crystal [9].

Remarkably, the authors referred the fast decay components to defects on the particle surfaces. TSL measurements have also shown the presence of defect states in the particles. There are, in fact, a lot of works dedicated to thermoluminescence in the undoped or europium doped LiCaAlF$_6$ single crystals [10,11-13]. All of them indicate the presence of irradiation-induced defects. However, the origin of these defects is still obscure. If they were paramagnetic, the method of electron paramagnetic resonance (EPR) could be used. Nonetheless, there is very tiny amount of papers reporting EPR studies. Mostly, they are concerned with dopants like Yb$^{3+}$ [14], Cr$^{3+}$ [15], Gd$^{3+}$ [16, 17], Fe$^{3+}$ [18] and $S$ spin state ions in general [19]. To the best of our knowledge, there is only one work discussing theoretical aspects of the bandgap, defects and activators [20]. As far as we know, there is no articles affording information about X-ray induced paramagnetic centers in the undoped LiCaAlF$_6$ as well as in the LiCaAlF$_6$:Eu single crystals. There is also lack of knowledge about paramagnetic Eu$^{2+}$ distribution and incorporation in the LiCaAlF$_6$ host especially taking into account the effect of sodium co-doping. In general, Eu can substitute each of the tree cations in the LiCaAlF$_6$ lattice leading to different optical and scintillation properties.

The present work is therefore focused on the investigation of nominally pure and europium, and sodium co-doped LiCaAlF$_6$ single crystals. The aim is to characterize peculiarities of the charge trapping, energy transfer and activator ions distribution in this lithium calcium aluminum fluoride lattice. In particular, we wanted to understand the origin of charge traps, to gain insight into the Eu$^{2+}$/Eu$^{3+}$ incorporation into the LiCaAlF$_6$ host and to clarify the impact of the Na co-doping on trapping centers and europium distribution. To fulfil these tasks, electron paramagnetic resonance and thermally stimulated luminescence techniques were engaged in a correlated manner.

## 2. Experimental

The Eu-doped LiCaAlF$_6$ (LiCAF) samples were the same as reported in [2] previously. The samples were cylindric shape about 1.5 mm in diameter. The undoped samples were grown by

Czochralski technique described in [1]. They were parallelepiped-shape cut of the 2x2.5x6 mm$^3$ size. The crystals were x-ray diffraction oriented along crystallographic axes.

The thermally stimulated luminescence (TSL) and radioluminescence (RL) were measured using the Horiba FluoroCube Spectrofluorometer equipped with a Janis liquid nitrogen cryostat and TBX-04 (IBH) photomultiplier operating in the 200–800 nm spectral range. The spectral resolution of the monochromator was 8 nm. The RL spectra were calibrated to spectral efficiency of the spectrofluorometer. The LiCAF samples were deposited on a silver paste on the sample holder. The samples were irradiated at 77 K by a Seifert X-ray tube operated at 40 kV with a tungsten target. All TSL measurements were performed in the temperature range 77−700 K and heating rate 0.1 K/s. The RL measurements were performed at 300 K and 77 K by a Horiba Jobin-Yvon setup.

Electron paramagnetic resonance measurements were performed with a commercial Bruker X/Q-band E580 FT/CW ELEXSYS and EMX plus spectrometers in the X-band (9.4 GHz) at temperatures 5 – 295 K. Prior to EPR measurements, each sample was weighted and the EPR intensity was normalized on the 1 g single crystal weight. Computer simulations of EPR spectra were carried out in the "Easyspin 5.2.27 toolbox" program [21].

For EPR measurements, the samples were also irradiated with an ISO-DEBYEFLEX 3003 highly stabilized X-ray equipment (tungsten X-ray tube, 50 kV, 30 mA) was used. According to the calibration curves the delivered dose to a sample was 15 kGy.

## 3. Results and discussion

*3.1. X-ray induced centers in the undoped LiCaAlF$_6$*

In the undoped LiCAF samples grown by Czochralski technique, only EPR signals of the Cr$^{3+}$ ions were visible. They can be seen in Fig. 1 along with the corresponding angular dependence in the (ac) plane. In the (aa) plane, the spectrum is insensitive to the sample rotation due to tetragonal symmetry of the paramagnetic center.

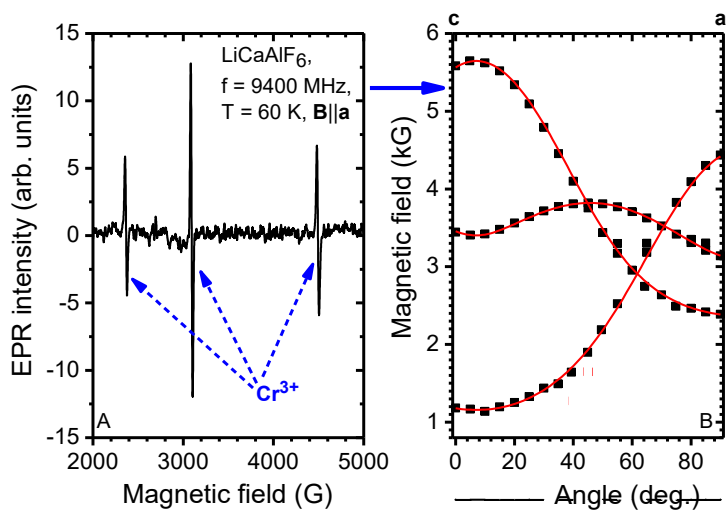

Fig. 1. $Cr^{3+}$ EPR spectrum measured along **a** crystal direction (A) at 60 K. The corresponding angular dependence in the (ac) plane is shown in panel (B). Dots represent experimental data whereas solid lines are calculated data.

To prove the correct guess of the impurity, the angular dependence has been fitted by the calculated one by using the following spin-Hamiltonian:

$$\hat{H}=\beta_e g \hat{S}_z H + \frac{1}{3}D\left(\hat{S}_z^2 - S(S+1)\right), \quad (1)$$

here $\beta_e$, $g$, $\hat{S}_z$, $H$, $D$ are, one by one, Bohr magneton, $g$ factor, $z$ component of the electron spin operator ($S=\frac{3}{2}$), magnetic field and axial zero field splitting (ZFS) constant [22]. The parameters of the fit are the following: $g=1.974\pm0.002$, $|D|=3100\pm20$ MHz $\approx 1033\times 10^{-4}$ cm$^{-1}$. These parameters are in good agreement with the data published in the previous work [23]: $g_\parallel = g_\perp = 1.974\pm0.002$, $D=-1010\pm20 \times 10^{-4}$ cm$^{-1}$ = 3028 MHz.

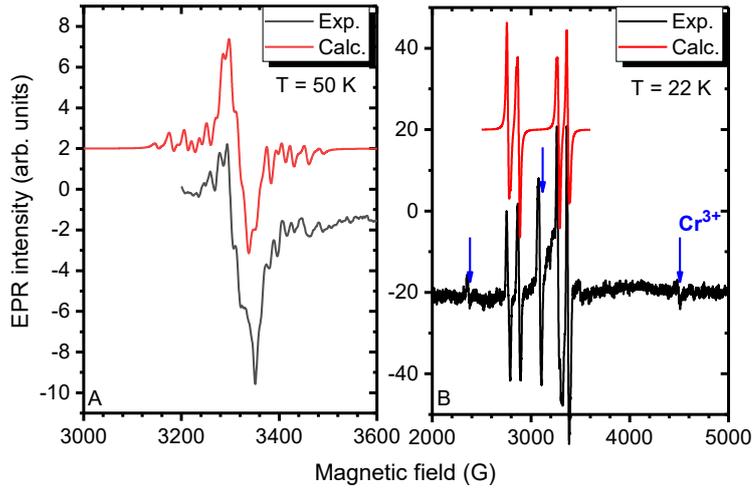

Fig. 2. X-ray irradiation induced EPR spectrum measured along **a** crystal direction of: (A) ClF$^-$ molecular ion at 50 K; (B) $F_2^- - F_2^-$ dimer measured at 22 K. Black lines represent experimental data whereas the red ones are calculated data. The $Cr^{3+}$ resonances are also indicated.

X-ray irradiation at room temperature results in the two new signals appearance shown in Fig. 2. The first one (panel A in Fig. 2) demonstrates a little angular dependence in the (ac) plane. It was deduced to be produced by the ClF$^-$ molecular ion, very similar to the well-known $V_k$ and H centers observed in halides [24-27], e.g., the $Cl_2^-$ in $Cs_2HfCl_6$ reported recently [26, 27]. In the present case, to prove the correct choice of the center's model, the experimental spectrum in Fig. 2A was fitted by the calculated one by using the two spin-Hamiltonians listed below:

$$\hat{H}_1 = \beta_e g_1 \hat{S}_z H + A_1\left(^{19}F\right)\hat{S}_z\hat{I}_z\left(^{19}F\right) + A_1\left(^{35,37}Cl\right)\hat{S}_z\hat{I}_z\left(^{35,37}Cl\right) + \sum_{i=1}^{4} A_{1i}\left(^{19}F\right)\hat{S}_z\hat{I}_z\left(^{19}F\right),$$

$$\hat{H}_2 = \beta_e g_2 \hat{S}_z H + A_2\left(^{19}F\right)\hat{S}_z\hat{I}_z\left(^{19}F\right) + A_2\left(^{35,37}Cl\right)\hat{S}_z\hat{I}_z\left(^{35,37}Cl\right) + \sum_{i=1}^{4} A_{2i}\left(^{19}F\right)\hat{S}_z\hat{I}_z\left(^{19}F\right),$$

(2)

The sum of two different spin Hamiltonians was needed to describe the spectrum which consists of contributions from two spin systems related to Cl and F ions. The parameters of fitting are: $g = g_1 = g_2 = 2.020 \pm 0.002$ and hyperfine constants $A_1\left(^{19}F\right) = 220 \pm 5$ MHz and $A_1\left(^{35,37}Cl\right) = 130 \pm 5$ MHz for the external magnetic field parallel with the line joining two anions. The $A_2\left(^{19}F\right) = 30 \pm 2$ MHz and $A_2\left(^{35,37}Cl\right) = 20 \pm 2$ MHz correspond to the spin system where the line joining two anions is perpendicular to the external magnetic field. The rest of the $A_{1i}\left(^{19}F\right)$ and $A_{2i}\left(^{19}F\right)$ are the contributions of the other fluorine nuclei of the regular octahedron, either $LiF_6$, $CaF_6$ or $AlF_6$ [28]. These constants were $A_{11}\left(^{19}F\right) = A_{12}\left(^{19}F\right) = 90 \pm 3$ MHz and $A_{13}\left(^{19}F\right) = A_{14}\left(^{19}F\right) = 80 \pm 3$ MHz for the first system and $A_{21}\left(^{19}F\right) = A_{22}\left(^{19}F\right) = 50 \pm 3$ MHz and $A_{23}\left(^{19}F\right) = A_{24}\left(^{19}F\right) = 30 \pm 3$ MHz for the second one. As one can see in Fig. 2A, the fit is almost perfect. The left edge of the experimental spectrum was strongly overlapped with other signals and thus was omitted in the present considerations. Since the spectrum demonstrated little of anisotropy, the fluorine anion of the $ClF^-$ molecular ion is expected to be the part of either $CaF_6$ or $AlF_6$ [28] octahedral complexes. The $LiF_6$ octahedron is more perturbed than the Ca and Al octahedrons. The signal (Fig. 2A) is very stable. It survives annealing at 480 K (it will be discussed further below). Therefore, the complementary chlorine anion is expected to be interstitial. The center model is shown in Fig. 3. There, the interstitial chlorine is connected to the fluorine in aluminum octahedron.

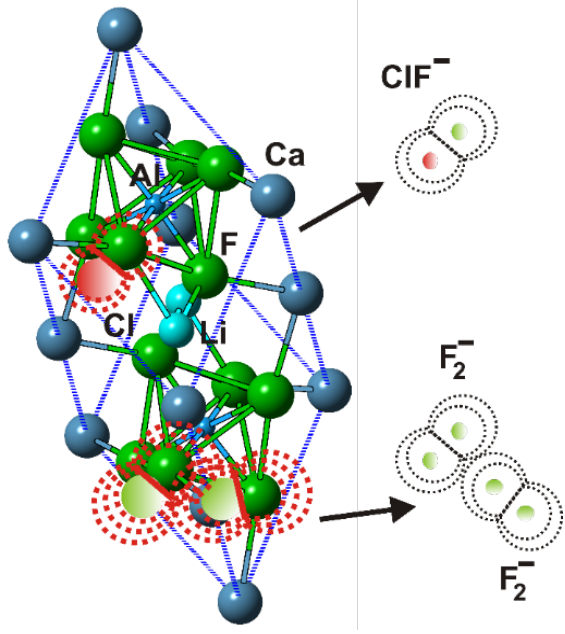

Fig. 3. The ClF$^-$ and $F_2^- - F_2^-$ center's models according to the EPR hyperfine data in Fig. 2 and spin Hamiltonian (2).

The second center created by the X-ray irradiation at RT issues the spectrum shown in Fig. 2B. After tentative analysis of its angular dependence it was possible to assume it to be a dimer created by two $F_2^-$ molecular anions exchange coupled as $F_2^- - F_2^-$. Assuming this model, the spectrum in Fig. 2B was fitted by using the following spin Hamiltonian:

$$\hat{H} = \beta_e \left( \hat{S}_{z1} g_1 + \hat{S}_{z2} g_2 \right) H + A_1^1 \left(^{19}F\right) \hat{S}_{z1} \hat{I}_{z1} \left(^{19}F\right) + A_2^1 \left(^{19}F\right) \hat{S}_{z2} \hat{I}_{z2} \left(^{19}F\right) +$$
$$+ A_3^1 \left(^{19}F\right) \hat{S}_{z1} \hat{I}_{z2} \left(^{19}F\right) + A_4^1 \left(^{19}F\right) \hat{S}_{z2} \hat{I}_{z1} \left(^{19}F\right) + A_1^2 \left(^{19}F\right) \hat{S}_{z1} \hat{I}_{z1} \left(^{19}F\right) + \quad , \quad (3)$$
$$+ A_2^2 \left(^{19}F\right) \hat{S}_{z2} \hat{I}_{z2} \left(^{19}F\right) + A_3^2 \left(^{19}F\right) \hat{S}_{z1} \hat{I}_{z2} \left(^{19}F\right) + A_4^2 \left(^{19}F\right) \hat{S}_{z2} \hat{I}_{z1} \left(^{19}F\right) + J_{zz} \hat{S}_{z1} \hat{S}_{z2}$$

here $A_i^1$ and $A_i^2$ ($i$ = 1-4) are the hyperfine constants of the $^{19}$F nuclei belonging to each of two $F_2^-$ molecular ions, 1 and 2 superscripts, in the $F_2^- - F_2^-$ dimer, respectively. For the $F_2^- - F_2^-$ spectrum simulation, the following parameters were used: $g_1 = g_2 = 2.173 \pm 0.003$, $A_1^1 = A_2^2 = 300 \pm 10$ MHz and $A_2^1 = A_3^1 = A_4^1 = A_1^2 = A_3^2 = A_4^2 = 30 \pm 3$ MHz. The exchange constant which includes all possible mechanisms of exchange along with the dipolar coupling was $J_{zz} = 3100 \pm 50$ MHz. It can be seen in Fig. 2B that the fit is perfect. This center also demonstrates advanced thermal stability surviving up to room temperature. Therefore, it was concluded to be a dimer of two H centers, with interstitial fluorine anion. This is not too rare event. This kind of centers has been studied already in halides, see e.g., [24]. The center model is shown in Fig. 3 as well.

Both new centers have their limits of thermal stability. To study them, the method of pulse annealing in air was used. The corresponding EPR spectra were measured at the reference temperature point, 50 K for the ClF$^-$ and 22 K for the $F_2^- - F_2^-$, whereas the annealing temperature was raised up each cycle of annealing, until the signals disappeared completely. Remarkably, a new signal (EC in Fig. 4) appeared as it is demonstrated in Fig. 4 for the 50 K reference temperature during this process. It has g factor $g = 2.0023$, the free electron value, and it is getting stronger continuously upon the annealing temperature. Its linewidth was approximately 6 G.

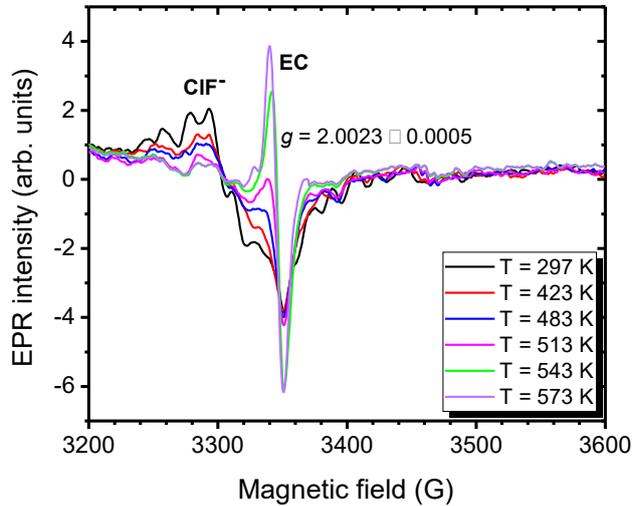

Fig. 4. Annealing data for the undoped LiCAF sample at the temperatures given in a legend. The decay of the the ClF$^-$ center EPR intensity and appearance with the consequent increase of the electron-like signal at the free electron g factor $g = 2.0023$ is shown. The spectra have been measured at 50 K. EC stands for an electron-like center.

The obtained annealing temperature dependencies of the EPR signals are shown in Fig. 5 along with the TSL glow curve measured in the same crystal.

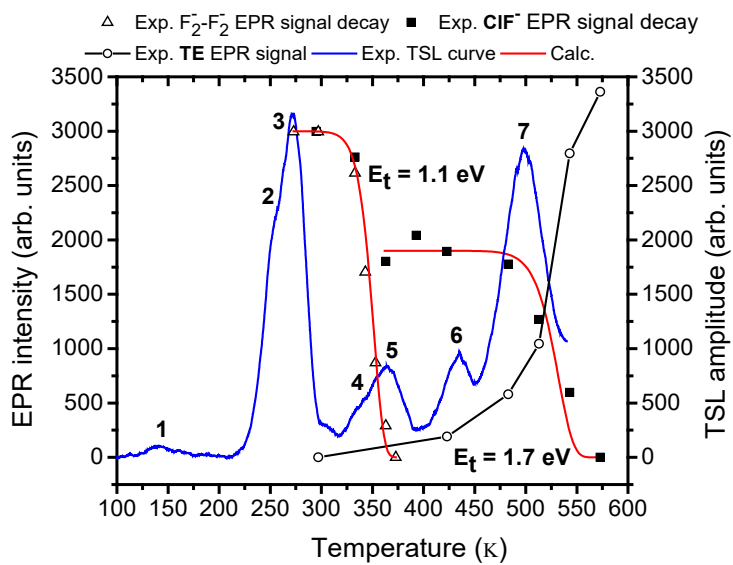

Fig. 5. Decay and increase of the EPR signals (mentioned in a legend) shown along with TSL glow curve. The correspondence between the decay of the $F_2^-$-$F_2^-$ and ClF$^-$ EPR signals and TSL peaks 4,5 and 7, respectively, can be observed. 1-7 enumerate TSL peaks.

The EC signal did not exist before the annealing (Fig. 4). It is neither connected to the X-ray irradiation or some charge re-trapping processes because even in the sample which has not been exposed to any kind of irradiation, the EC signal appears in the same way as it is shown in Fig. 4. It demonstrates the same intensity dependence on the annealing temperature as shown in Fig. 5 for the X-ray irradiated sample. The EC center is thus a signal produced by dangling bonds on the crystal surface. Interaction of the surface with oxygen in air creates more paramagnetic specimens of that kind resulting in the increased EC EPR signal.

The correlation between the $ClF^-$ EPR signal decay curve and the glow peak 7 (490 K) and the $F_2^- - F_2^-$ signal decay curve and the glow peaks 4, 5 (360 K) can be found. Note, that the 340, 420 and 490 K peaks were observed previously in undoped and $LiCaAlF_6$:Eu samples [10]. Small temperature shift of the EPR decay curves with respect to the glow peaks is due to different setups used in EPR and TSL experiments and different ways of temperature measurement. The EPR decay curves were fitted with the recursive expression [27, 29]:

$$I_{i+1} = I_i \exp\left(-f_0 t \exp\left(-E_t / k_B T_i\right)\right), \qquad (4)$$

where $I_i$ is the EPR intensity prior to the $i$-th cycle of annealing at the chosen temperature $T_i$ for $t = 4$ minutes. The following $E_t(ClF^-) = 1.7 \pm 0.1$ eV and $E_t(F_2^- - F_2^-) = 1.1 \pm 0.1$ eV trap depths, and $f_0(ClF^-) = f_0(F_2^- - F_2^-) \sim 10^{13}$ s$^{-1}$ frequency factors were determined from the fit.

*3.2. Europium doped LiCaAlF$_6$. The effect of Na co-doping*

EPR spectra measured in the Eu-doped and Na co-doped LiCAF single crystals are shown in Fig. 6. Magnified, the same spectra are demonstrated in Fig. 7.

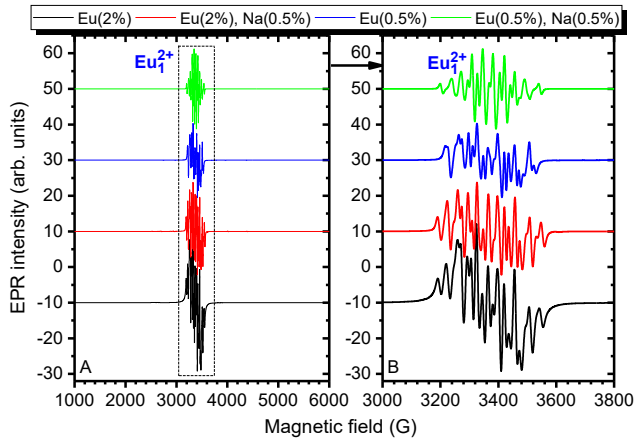

Fig. 6. Eu$^{2+}$ EPR spectra in the LiCAF:Eu,Na single crystals as measured (A) and zoomed in (B). The dominating cubic symmetry Eu$^{2+}$ spectrum ($Eu_1^{2+}$) is clearly visible. Europium and sodium contents are given in a legend.

One can immediately see that the dominating is the cubic $Eu^{2+}$ spectrum at $g$ factor $g = 1.99$ typical for $Eu^{2+}$ ions [30, 31], $Eu_1^{2+}$. This spectrum does not markedly depends on crystal orientation supporting cubic symmetry of the $Eu_1^{2+}$ center. This kind of spectrum has been observed recently in the europium-doped $NaLuS_2$ and sodium containing mixed ternary sulphides [32, 33]. Its intensity is insensitive to the presence of the Na co-dopant. The second, trigonal symmetry $Eu^{2+}$ spectrum, $Eu_2^{2+}$, is approximately 20 times weaker than the cubic one, can be observed only in the magnified spectra shown in Fig. 7.

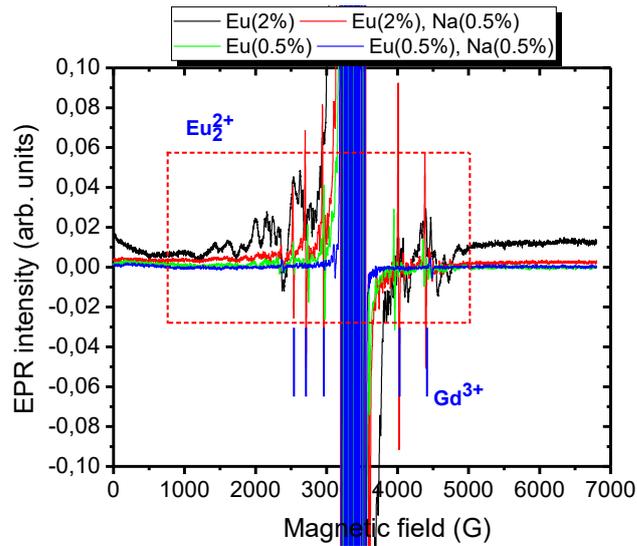

Fig. 7. Magnified $Eu^{2+}$ EPR spectra in the LiCAF:Eu,Na single crystals. The weak trigonal $Eu^{2+}$ spectrum ($Eu_2^{2+}$) is demonstrated. Europium and sodium contents are given in a legend.

The $Eu_2^{2+}$ spectrum intensity strongly depends on the sodium content. The trigonal europium signal is getting significantly weaker when the Na is present in large concentration. $Na^+$ is more suitable in the $Li^+$ site than $Eu^{2+}$ because lithium and sodium belong to the same group of alkali metals. Their ionic radii in the six-fold coordination are $r_{Na} = 1.02$ Å and $r_{Li} = 0.76$ Å, so the difference is $r_{Na} - r_{Li} = 0.26$ Å [34]. The ionic radius of a $Eu^{2+}$ ion is $r_{Eu} = 1.17$ Å in the same coordination and thus, the $r_{Eu} - r_{Li} = 0.41$ Å difference is even larger. Calcium ionic radius in the six-fold coordination is $r_{Ca} = 1$ Å [34], the $r_{Eu} - r_{Ca} = 0.17$ Å << 0.41 Å. Therefore, the dominating amount of $Eu^{2+}$, the $Eu_1^{2+}$ is expected to substitute for Ca, whereas much lower concentration of $Eu^{2+}$, the $Eu_2^{2+}$ presumably substitutes for Li, which site is likely to be occupied by Na as well. The LiCAF:Eu,Na contains also small traces of $Gd^{3+}$. Both $Eu^{2+}$ spectra reveal no sign of changes after the X-ray irradiation. However, the $F_2^- - F_2^-$ signal is clearly visible (only two leftmost resonance lines, see Fig. 2B) even though it is mixed with the $Eu_1^{2+}$. It is demonstrated in Fig. 8, on example of the LiCAF:Eu(0.5%) crystal.

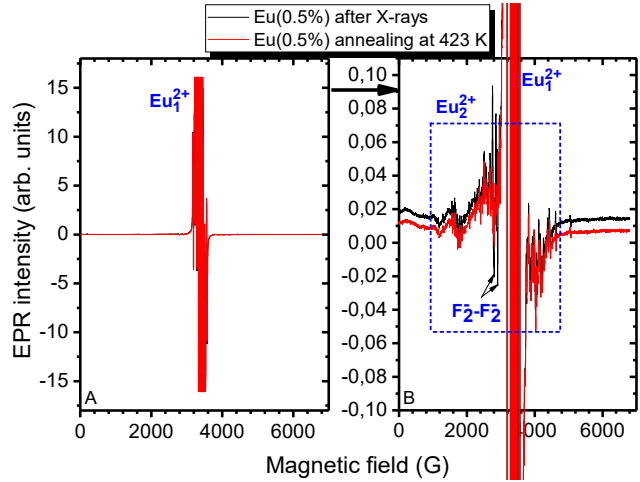

Fig. 8. (A) $Eu^{2+}$ EPR spectrum in the LiCAF:Eu(0.5%) single crystal measured at 30 K before and after exposure to the X-ray irradiation. (B) The same spectra zoomed in. The $F_2^- - F_2^-$ EPR signal mixed up with the $Eu_1^{2+}$ and $Eu_2^{2+}$ spectra is indicated as well.

If the ClF$^-$ EPR signal existed in the LiCAF:Eu(0.5%) and LiCAF:Eu(0.5%),Na(0.5%), it would be too strongly overlapped with the dominating $Eu_1^{2+}$ EPR signal to be figured out.

The europium-doped crystals have been also studied by RL and TSL techniques. The RL spectra measured at 295 and 77 K are shown in Fig. 9.

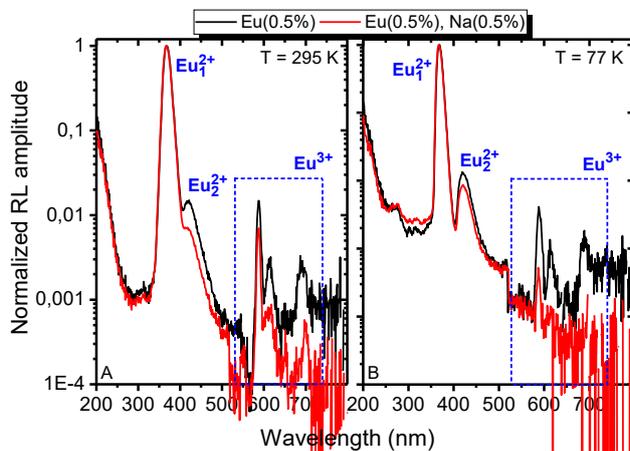

Fig. 9. RL spectra measured in the LiCAF:Eu(0.5%) and LiCAF:Eu(0.5%),Na(0.5%) at 295 K (A) and 77 K (B). The $Eu_1^{2+}$, $Eu_2^{2+}$ and $Eu^{3+}$ bands are indicated.

The strongest band peaking at approximately 380 nm was insensitive to the sodium presence. It was referred to the $Eu_1^{2+}$ center, whereas the much weaker one (by about two orders of magnitude) at 420 nm should be issued by the $Eu_2^{2+}$. This assumption was made considering drop of the band intensity upon the sodium co-doping correlated to the changes in the $Eu_2^{2+}$ EPR signal (Fig. 7). The same supposition has been made in the previous work [10]. The $Eu^{3+}$ lines were recognized basing on the data obtained in the LiCaAlF$_6$:Eu [35]. Since the Eu$^{3+}$ transitions are of the inner 4f-4f type, shielded from the influence of the local crystal field, their spectral positions remain approximately the same in all hosts. The Eu$^{3+}$ bands were very weak as compared to the Eu$_1^{2+}$ band proving that the europium prefers to stay in the 2+ charge state in the host. RL spectra measured at both 295 and 77 K demonstrate the same Eu$_1^{2+}$ : Eu$_2^{2+}$ : Eu$^{3+}$ intensity ratio. No artefacts were discovered by lowering the temperature from the room to liquid nitrogen one.

To check the possible changes in the defects states and their distribution compared to the undoped sample, TSL glow curves were measured in the LiCAF:Eu,Na single crystals. They are shown in Fig. 10 on example of the LiCAF:Eu(0.5%), LiCAF:Eu(0.5%),Na(0.5%) and undoped LiCAF samples.

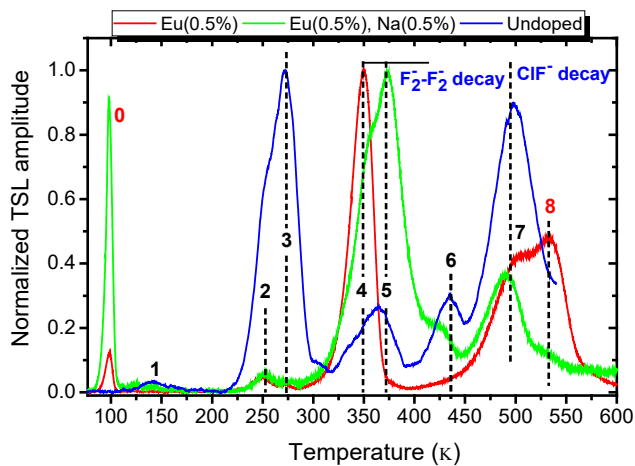

Fig. 10. TSL glow curves measured in the LiCAF:Eu(0.5%), LiCAF:Eu(0.5%),Na(0.5%) and undoped LiCAF samples. Peaks 1-7 have already been demonstrated in Fig. 5 in the undoped sample whereas the peaks 0 and 8 are new.

The peaks 1-7 have already been detected in the undoped sample (Fig. 5). They are also present in the LiCAF:Eu(0.5%) and LiCAF:Eu(0.5%),Na(0.5%). The peak 1 is too weak to compare, however, the peaks 2, 3 being strong in the undoped sample became at least one order of magnitude weaker in both the LiCAF:Eu(0.5%) and LiCAF:Eu(0.5%),Na(0.5%) samples. Oppositely, the peak 4 is strongly increased in the LiCAF:Eu(0.5%). On the other hand, the peak 5 seems to be completely reduced. In the LiCAF:Eu(0.5%),Na(0.5%) sample, both peaks 4, 5 have raised up significantly. Note, that namely these peaks were related to the $F_2^- - F_2^-$ EPR signal decay in the undoped sample (see Fig. 5). The peak 6 is completely lost in the LiCAF:Eu(0.5%) and partially in the LiCAF:Eu(0.5%),Na(0.5%) samples. The peak 7 related to the ClF$^-$ EPR signal decay curve (Fig. 5) is decreased in both LiCAF:Eu(0.5%) and LiCAF:Eu(0.5%),Na(0.5%) samples. Except these 7 peaks, two new bands appeared in the LiCAF:Eu(0.5%) and LiCAF:Eu(0.5%),Na(0.5%) glow curves indicated as 0 (100 K) and 8 (530 K) in Fig. 10. The peak 0 is approximately 6 times weaker in the LiCAF:Eu(0.5%) as compared to the LiCAF:Eu(0.5%),Na(0.5%). The peak 8 is approximately 4 times weaker in the LiCAF:Eu(0.5%),Na(0.5%) as compared to the LiCAF:Eu(0.5%). All the peaks except the 4, 5 and 7 have no obvious connection to the paramagnetic impurities and thus their origin is unclear. The discussed changes in the glow peaks intensities upon the europium doping and sodium co-doping make evidence for the strong affectation of the already observed unstable defects or even creation of the new ones. In particular, the peak 0 may be connected with the Li site occupation. In the LiCAF:Eu(0.5%), some small part of the overall Eu is stabilized as the $Eu_2^{2+}$ center, Eu at Li site, so the peak 0 appeared. The addition of sodium in the LiCAF:Eu(0.5%),Na(0.5%) with the Na placed at Li makes it even stronger. Since the peak 4 exists in the LiCAF:Eu(0.5%) and the $F_2^- - F_2^-$ EPR signal too (Fig. 8B) while the peak 5 is absent, namely the peak 4 should be connected with the $F_2^- - F_2^-$ decay. Europium also serves as inhibitor of the trap states giving the rise of peaks 2,3, most probably, because of the charge state changes, so the trap responsible for those peaks prior to the europium doping is now deactivated. It could be possible, because of the $Eu_2^{2+}$ center existence in the Li site. This means that extra positive charge is necessary for compensation.

**Conclusion**

Several single crystal samples of undoped and Eu, Na co-doped LiCaAlF$_6$ with different europium and sodium content were studied by EPR, RL and TSL. EPR measurements have shown the presence of only Cr$^{3+}$ ions at very low level before exposure to X-ray irradiation in the undoped sample. After the irradiation at liquid nitrogen temperature, the Cr$^{3+}$ EPR signals remained the same whereas two new spectra belonging to hole-like centers were discovered. These hole centers were molecular ions of the form: ClF$^-$ and $F_2^- - F_2^-$ dimer. The decay curves of the corresponding EPR signals intensity as a function of annealing temperature were obtained. The $F_2^- - F_2^-$ dimer and ClF$^-$ trap depths and frequency factors, $E_t\left(ClF^-\right) = 1.7 \pm 0.1$ eV and $E_t\left(F_2^- - F_2^-\right) = 1.1 \pm 0.1$ eV and $f_0\left(ClF^-\right) = f_0\left(F_2^- - F_2^-\right) \sim 10^{13}$ s$^{-1}$ were determined. A new electron-like EPR signal became visible and its intensity was rising upon the annealing temperature (in air). The same dependence was also observed even in the sample which has never experienced to any kind of irradiation. Therefore, this signal was referred to the dangling bonds on the sample surface and not to some re-trapping processes. The number of these paramagnetic species was getting larger because of the intense interaction of the sample surface with oxygen in air at elevated temperatures. The $F_2^- - F_2^-$ dimer and ClF$^-$ decay curves were correlated to two TSL glow peaks having maxima at approximately 360 K ($F_2^- - F_2^-$ dimer) and 490 K (ClF$^-$). In total, however, the TSL glow curve demonstrated 7 peaks in the undoped sample.

EPR spectra of the europium doped samples demonstrated two different Eu$^{2+}$ signals, the cubic $Eu_1^{2+} = Eu_{Ca}^{2+}$ and trigonal $Eu_2^{2+} = Eu_{Li}^{2+}$ symmetry. The first one is strongly dominating. Spin-Hamiltonian parameters have been determined for both these centers. The Eu$^{2+}$ centers are insensitive to X-ray irradiation. The $F_2^- - F_2^-$ dimer EPR signal can be observed after the irradiation, same as in the undoped sample. Sodium admixture results in drop of the $Eu_2^{2+} = Eu_{Li}^{2+}$ EPR signal. Sodium which is smaller than $Eu^{2+}$ belongs to the same alkali metals as lithium and thus it is preferable for the Li site. This allowed to attribute the strongest 380 nm band in RL spectra to the $Eu_1^{2+} = Eu_{Ca}^{2+}$ and the very weak 420 nm one to the $Eu_2^{2+} = Eu_{Li}^{2+}$ because the correlation between EPR and RL intensities of the europium centers was observed. The $Eu^{3+}$ bands weaker than the $Eu^{2+}$ ones were also detected in RL spectra. All this leads to the conclusion that the preferable charge state of europium is 2+ in the $LiCaAlF_6$.

TSL glow curves measured in the $LiCaAlF_6$: Eu,Na demonstrate redistribution of intensity over the peaks observed in the undoped sample. In particular, the 360 K peak related to the thermal decay of the $F_2^- - F_2^-$ dimer is getting larger. The 490 K peak is lower than in the undoped sample. Besides, the two other peaks appeared at 100 K and 530 K. The 100 K one had a tendency to increase upon dopant concentration, either europium or sodium. All this creates evidence for the strong influence of europium and sodium dopants on the defect states in the $LiCaAlF_6$.

**Acknowledgements**

The authors gratefully acknowledge the financial support of the Czech Science Foundation (project No. 17-09933S) and the Ministry of Education, Youth and Sports of Czech Republic (project No. CZ.02.1.01/0.0/0.0/16_013/0001406).